\documentclass{article}
\usepackage{arxiv}
\usepackage[numbers,sort&compress]{natbib}
\usepackage[utf8]{inputenc} 
\usepackage[T1]{fontenc}    
\usepackage{ dsfont }
\usepackage{hyperref}       
\usepackage{url}
\usepackage{textcomp}
\usepackage{ gensymb }
\usepackage{amsmath}
\usepackage[ruled,vlined]{algorithm2e}
\usepackage{algorithmic}
\usepackage{booktabs}       
\usepackage{amsfonts}       
\usepackage{nicefrac}       
\usepackage{microtype}      
\usepackage{lipsum}
\usepackage{float}
\usepackage{graphicx}
\usepackage{multirow}
\usepackage{array}
\usepackage{comment}
\usepackage{longtable}
\usepackage{ragged2e}
\usepackage{subfig}

\title{Defect Localization Using Region of Interest and Histogram-Based Enhancement Approaches in 3D-Printing}

\author{
  Md Manjurul Ahsan \\
  Department of Industrial and Systems Engineering\\
  University of Oklahoma\\
  Norman, Oklahoma-73071 \\
  \texttt{ahsan@ou.edu} \\
   \And
 Shivakumar Raman \\
  Department of Industrial and Systems Engineering\\
  University of Oklahoma\\
  Norman, Oklahoma-73071\\
  \texttt{raman@ou.edu}\\
  \And
 Zahed Siddique \\
  School of Aerospace and Mechanical Engineering\\
  University of Oklahoma\\
  Norman, Oklahoma-73019\\
  \texttt{zsiddique@ou.edu}} 

\begin{document}
\maketitle
\begin{abstract}
Additive manufacturing (AM), particularly 3D printing, has revolutionized the production of complex structures across various industries. However, ensuring quality and detecting defects in 3D-printed objects remain significant challenges. This study focuses on improving defect detection in 3D-printed cylinders by integrating novel pre-processing techniques such as Region of Interest (ROI) selection, Histogram Equalization (HE), and Details Enhancer (DE) with Convolutional Neural Networks (CNNs), specifically the modified VGG16 model. The approaches, ROIN, ROIHEN, and ROIHEDEN, demonstrated promising results, with the best model achieving an accuracy of 1.00 and an F1-score of 1.00 on the test set. The study also explored the models' interpretability through Local Interpretable Model-Agnostic Explanations and Gradient-weighted Class Activation Mapping, enhancing the understanding of the decision-making process. Furthermore, the modified VGG16 model showed superior computational efficiency with 30713M FLOPs and 15M parameters, the lowest among the compared models. These findings underscore the significance of tailored pre-processing and CNNs in enhancing defect detection in AM, offering a pathway to improve manufacturing precision and efficiency. This research not only contributes to the advancement of 3D printing technology but also highlights the potential of integrating machine learning with AM for superior quality control.

\end{abstract}


\maketitle

\section{Introduction}

Since it appeared in the 1980s, additive manufacturing (AM) has changed how we make three-dimensional (3D) objects. 3D printing, a key AM technique, builds objects by layering materials one after the other. This approach has gained a lot of attention for its ability to make complex shapes that traditional methods cannot~\cite{article,bhusnure20163d,javan2018prototype,zou20233d}. The technique has greatly broadened the types of materials we can print, including polymers, metals, ceramics, composites, and biological materials~\cite{khorasani2022review,ramezani20234d}. Starting as a rapid prototyping tool, 3D printing now extends across many fields such as biomedical, aerospace, automotive, and consumer products, demonstrating its wide use and transformative effect~\cite{nazir2023multi,siddiqui2023emerging}. The ability to efficiently fabricate custom components on demand with reduced waste has been pivotal in manufacturing, supply chain management, and biomedical fields~\cite{park20223d,shahrubudin2020challenges}. However, challenges remain in terms of material diversity, build size, printing speed, output quality, and cost-effectiveness~\cite{dabbagh20223d,kantaros2021manufacturing}. Addressing these issues through advancements in printing technology, materials science, design software, machine learning, and systems integration is crucial for harnessing the full potential of 3D printing~\cite{lalegani2021overview,ansari2022layer}. Progress in these interdisciplinary areas is essential for the continued evolution of 3D printing technologies. 

\begin{figure*}[ht]
    \centering
    \includegraphics[width=.9\textwidth]{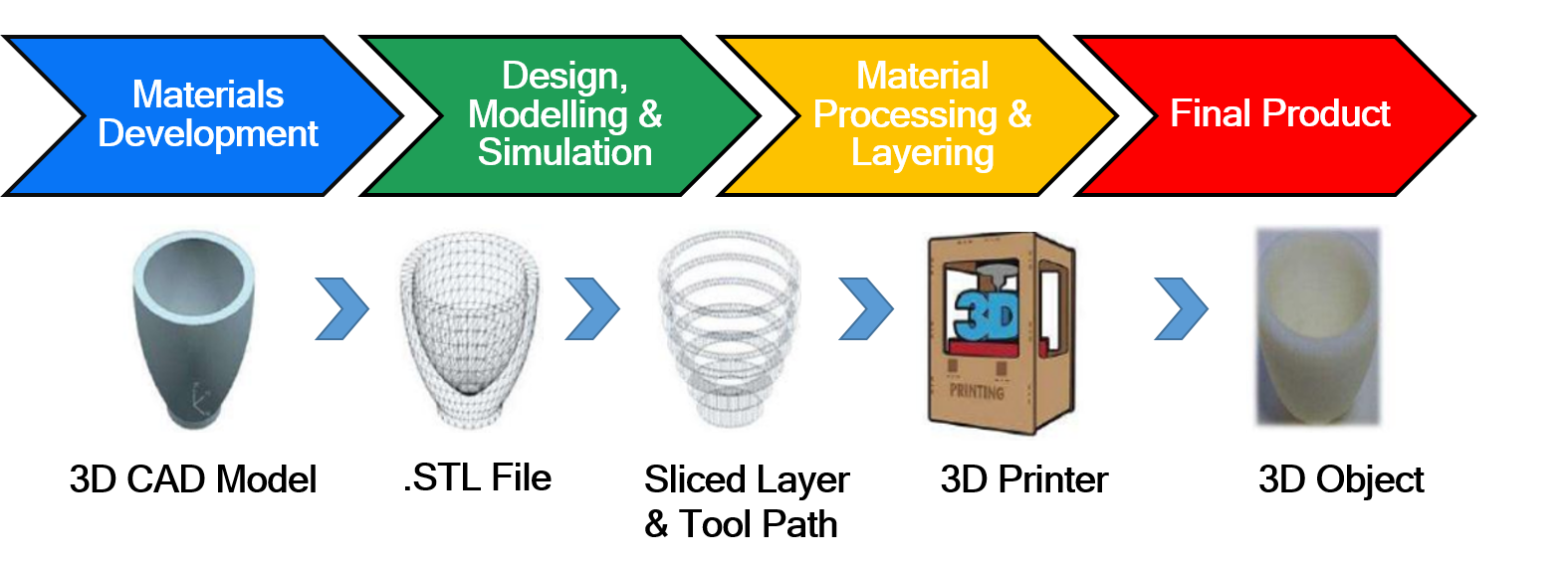}
    \caption{A standard 3D printing process flow diagram, illustrating the progression from materials development to the final product~\cite{article}.}
    \label{fig:3dprintp}
\end{figure*}

\textbf{Figure}~\ref{fig:3dprintp} provides a general overview of the typical 3D-printing process, which includes several steps such as materials development, design, modeling, simulations, materials processing, and layering as well.

In recent years, machine learning (ML) approaches have demonstrated considerable promise in surmounting myriad challenges confronting 3D printing technology~\cite{belei2022fused,haleem20213d}. A salient application entails predicting and averting printing defects by training ML models on expansive datasets, thereby enabling the identification of patterns and prediction of defects including warping, shrinkage, and layer separation. Such capabilities facilitate process optimization and enhancement of printed part quality~\cite{mazzanti2019fdm}. Additionally, ML enables optimization of printing parameters such as speed, temperature, and layer height to curtail printing duration and energy consumption while preserving quality standards~\cite{paraskevoudis2020real}. Moreover, ML can assist in developing novel materials tailored for 3D printing, consequently expanding the gamut of printable materials~\cite{ansari2022layer}. Enhancing the accuracy of 3D scanning and modeling, an indispensable precursor in 3D printing, also falls within the purview of ML~\cite{haleem20213d}. In summary, ML harbors immense potential to ameliorate 3D printing capabilities and conquer existing challenges~\cite{belei2022fused}.

Several methods have been utilized to identify defects in 3D-printed components. Visual inspection, whereby printed parts are visually examined for defects including warping, cracking, or uneven surfaces, represents one common yet subjective and time-consuming approach, rendering it impractical for large-scale manufacturing~\cite{wang2022role}. Microscopy techniques such as scanning electron microscopy (SEM) or optical microscopy enable examining the microstructure of printed parts to identify defects like porosity or voids at high resolution. However, microscopy is often destructive and limited by component size~\cite{demeneghi2021size}. 

Non-destructive testing (NDT) methods, including X-ray computed tomography (CT) or ultrasound, provide detailed internal images to detect defects such as voids or delamination. Nevertheless, NDT techniques are expensive and time-consuming, restricting their scalability for high-volume manufacturing~\cite{sun2022review}. Overcoming the limitations of current defect detection methods requires developing rapid, accurate, and cost-effective approaches scalable for industrial 3D-printing.

Inspecting 3D printed components visually for defects such as warping, cracking, or uneven surfaces represents a common yet subjective and time-intensive approach, rendering it impractical for large-scale manufacturing~\cite{wang2022role}. Microscopy techniques, including scanning electron microscopy (SEM) and optical microscopy, enable examining a part's microstructure to identify defects like porosity at high resolution but remain destructive and constrained by part size~\cite{demeneghi2021size}. Non-destructive testing (NDT) methods such as X-ray computed tomography (CT) and ultrasound furnish detailed internal images revealing voids or delamination, though proving expensive and having limited scalability~\cite{sun2022review}.

Surmounting present limitations in defect detection necessitates developing novel approaches that are rapid, accurate, cost-effective, and scalable. Targeted research towards integrating automation, machine learning, and process data analytics appears promising for devising next-generation defect detection paradigms suitable for industrial 3D printing applications~\cite{ansari2022layer,belei2022fused,haleem20213d}. Nevertheless, significant interdisciplinary advances across manufacturing, materials, and computer science are requisite to realize facile, large-scale quality assurance for 3D printing.

However, certain limitations are associated with applying Transfer Learning (TL) and Convolutional Neural Network (CNN)-based methods in identifying defects in images of 3D printed cylinders. These limitations include:
\begin{itemize}
    \item The effectiveness of CNNs in detecting defects in 3D printing relies on the quality of the training dataset, which necessitates careful curation to ensure its representation of the various types of defects encountered in practical scenarios~\cite{wen2021application}.

    \item Detecting 3D printing defects, particularly internal irregularities like voids or porosity that may not be visible on the printed part's surface presents a significant challenge~\cite{dhakal2023impact}.
    \item TL requires the availability of a pre-trained model relevant to the specific application, and it may be challenging to find pre-trained models suitable for 3D printing defect detection~\cite{saeed2019automatic}.
    \item Interpreting the results produced by CNNs and TL can be challenging, especially for complex models, making it difficult to determine which features are most important for defect detection~\cite{liu2021defect}.
\end{itemize}
This study focuses on classifying 3D-printed cylindrical objects as either defective or non-defective using image classification techniques. Employing a dataset that includes both balanced and imbalanced samples from the Smart Materials and Intelligent Systems (SMIS) Laboratory at the University of Oklahoma, it addresses the significant challenge of data imbalance, which can affect classification accuracy and introduce biases.

The primary aim is twofold: to achieve precise and reliable object classification while ensuring the outcomes are interpretable and explainable. This effort seeks to make the decision-making process transparent, highlighting the rationale behind classifications and the significant features influencing these decisions.

By overcoming these challenges, the research contributes to the development of dependable and clear image classification methods for detecting defects in 3D printing, thereby improving the process's quality and efficiency. These advancements are expected to spur progress in 3D printing, enhancing manufacturing precision and productivity.
The technical contribution of this study can be summarized as follows:

\begin{itemize}
    \item This study proposes an update to an existing model, explicitly focusing on enhancing the modified VGG16 model. It assesses the model's performance against established Convolutional Neural Network (CNN) models, particularly evaluating its computational efficiency.
    \item By introducing measures to localize defect regions within 3D-printed cylinders, this research endeavors to enhance the quality of 3D-printed products and simultaneously reduce production time and costs. 
    \item The proposed approach presents an effective solution for industries adopting economically viable AI-based Additive Manufacturing (AM) techniques. 
\item Offers valuable insights into the integration of image processing techniques and Transfer Learning (TL)-based models for the precise identification of defect regions within 3D-printed products.
\item The findings of this study hold significance for a broad spectrum of stakeholders, including researchers, practitioners, and industries. It equips them with knowledge to optimize production processes, ultimately enhancing their competitiveness within the market.
\item The proposed approach encompasses three essential pre-processing steps: Region of Interest (ROI) selection, Histogram Equalization (HE), and Details Enhancer (DE).
\item This study addresses the challenge of localizing specific defect regions within 3D-printed cylinder images, a limitation encountered even by highly accurate TL-based CNN models.
\item The research underscores the critical role of pre-processing steps, particularly in identifying defect regions within cylinder images, to bolster CNN model performance.
\item Employing LIME and Grad-CAM techniques, this study enhances the interpretability of model predictions. It facilitates the identification of pivotal features contributing to model decisions and offers insights into the model's decision-making process.
\item The proposed approaches, including ROIN, ROIHEN, and ROIHEDEN, demonstrate promise in detecting defects in 3D-printed cylinder images. These models achieved high accuracy, precision, recall, F1-score values, and sensitivity, thus presenting viable solutions for defect detection in such images.
\end{itemize}
The remainder of the paper is structured as follows: Section~\ref{mot} discusses the motivation, Section~\ref{proa} describes the proposed approaches, Section~\ref{observation} presents the results, Section~\ref{discussion} offers a discussion of the findings, and Section~\ref{con} concludes the paper and suggests future research directions. 
\section{Related Works}\label{rl}
In recent years, numerous Deep Learning (DL) researchers have proposed Machine Learning (ML) methods to enhance 3D printing efficiency. For example, Saeed et al. (2019) introduced an autonomous post-processing approach using CNN and DFF-NN algorithms to detect defects and estimate thermogram depth, offering real-time, automated post-processing for production lines. They also comprehensively reviewed CNN's application to IR thermograms, presenting comparative results of various CNN models and object detection frameworks. They demonstrated its potential for speeding up inspections and integrating with production facility quality control systems~\cite{saeed2019automatic}.

Jia et al. (2022) presented a fabric defect detection system that utilizes transfer learning and an improved Faster R-CNN to accurately address the challenges of detecting small target defects. Their system achieved superior accuracy and convergence compared to existing models by incorporating pre-trained weights from the Imagenet dataset, leveraging ResNet50 and ROI Align for feature extraction, and using RPN and FPN in combination with different IoU thresholds for sample classification~\cite{jia2022fabric}.

Chen et al. (2023) introduced a deep learning method for detecting low-contrast defects in 3D-printed ceramic curved surface parts. This approach employed a blurry inpainting network model and a multi-scale detail contrast enhancement algorithm to enhance image quality and feature information. Using the ECANet-Mobilenet SSD network model, the method achieved high accuracy in detecting crack and bulge defects, contributing to intelligent defect detection in the advanced ceramic industry, albeit focusing on specific defect types and datasets~\cite{chen2023defect}.

Li et al. (2022) provided an extensive survey on Deep Transfer Learning (DTL) techniques in Intelligent Fault Diagnosis (IFD). While the survey covered theoretical DTL background and recent IFD developments, it primarily offered recommendations for selecting DTL algorithms in practical applications and identifying future challenges. However, empirical results and experiments were lacking, and the focus remained on DTL-based IFD methods~\cite{li2023ifd}.

Ma et al. (2022) proposed a transitive transfer learning CNN ensemble framework for bearing surface defect detection and classification, addressing limited dataset challenges in deep learning. This framework achieved a high accuracy rate and met industrial online detection requirements. Although it lacked comparisons with other methods and focused on a specific defect type, it displayed promise for enhancing the efficiency and accuracy of bearing quality inspection in the industry~\cite{ma2022novel}.

The studies offered innovative additive manufacturing (AM) solutions and quality assurance. Lu et al. (2023) introduce a real-time defect detection and closed-loop adjustment system driven by deep learning tailored explicitly for carbon fiber reinforced polymer (CFRP) composites. While limited to CFRP composites, it effectively identifies and controls defects through process parameter adjustments. Westphal (2022) presents a novel machine-learning approach for quality assurance in Fused Deposition Modeling (FDM) AM processes, utilizing environmental sensor data and machine-learning algorithms. The findings indicate ML analyses as a more efficient alternative for discerning various 3D printing conditions, enhancing trust in industrial additive manufacturing. Caggiano et al. (2019) propose a machine learning approach, leveraging a bi-stream deep convolutional neural network (DCNN) to identify defects in Selective Laser Melting (SLM) of metal powders. Their work enables adaptive process control and quality assurance, showcasing promise for industrial applications in SLM quality assurance~\cite{lu2023deep, westphal2022machine, caggiano2019machine}.

Wang et al. (2020) provide an extensive review of machine learning (ML) applications in additive manufacturing (AM), exploring its potential across various domains. They delve into ML's capacity in material design, topology design, process parameter optimization, in-process monitoring, and product quality assessment and control within AM. The review underscores the importance of data security in AM and outlines future research directions. While current ML applications in AM primarily focus on processing-related aspects, the review anticipates a shift towards exploring novel materials and rational manufacturing~\cite{wang2020machine}.

In a complementary effort, Qi et al. (2019) conducted a review focusing on neural networks (NNs) within additive manufacturing (AM), particularly for complex pattern recognition and regression analysis. NNs offer a promising avenue to tackle the challenge of constructing and solving physical models concerning the process-structure-property-performance relationship in AM. The review spans various application scenarios, including design, in-situ monitoring, and quality assessment, while also addressing challenges in data collection and quality control with potential solutions. The authors emphasize the significant potential of synergizing AM and NNs to usher in agile manufacturing practices in the industry~\cite{qi2019applying}.

Kwon et al. (2020) utilized a deep neural network to classify selective laser melting melt-pool images based on six laser power labels, achieving satisfactory inference, even with images featuring blurred edges. The proposed neural network demonstrated a low classification failure rate (under 1.1\%) for 13,200 test images and outperformed simple calculations in monitoring melt-pool images. This classification model holds the potential for non-destructive inference of microstructure alterations or identifying defective products. However, potential limitations of deep neural networks in selective laser melting classification were not addressed in the study~\cite{kwon2020deep}.

Aquil et al. (2020) explored software defect prediction using machine learning techniques, including deep learning, ensembling, data mining, clustering, and classification. The study assessed various predictor models using 13 software defect datasets and consistently found high-accuracy predictions with ensembling techniques. However, it lacked specificity regarding the types of software defects predicted and dataset representativeness. The findings suggest the reliability of ensembling techniques for software defect prediction but underscore further research to identify optimal predictor models for diverse software systems~\cite{aquil2020predicting}.

Dogan et al. (2021) conducted a comprehensive literature review on the application of machine learning and data mining in manufacturing, categorizing solutions into scheduling, monitoring, quality, and failure domains. The paper highlighted the advantages of employing machine learning in manufacturing and presented statistical insights into the field's current state. It also explained the knowledge discovery steps in the databases (KDD) process for manufacturing applications but did not delve into the limitations and challenges of implementing machine learning in manufacturing~\cite{dogan2021systematic}.

Majeed et al. (2021) introduced the BD-SSAM framework, which integrates big data analytics, additive manufacturing, and sustainable smart manufacturing to support decision-making during the early stages of the product lifecycle. Implemented in the fabrication of AlSi10Mg alloy components using selective laser melting, the framework effectively controlled energy consumption and improved product quality. It promoted smart, sustainable manufacturing, emission reduction, and cleaner production. However, the study's scope was limited to the early product lifecycle stages and a specific material (AlSi10Mg alloy components with SLM). Further research is necessary to validate the framework across different materials and product lifecycle stages, although it presents a promising approach for smart, sustainable manufacturing in the additive manufacturing industry~\cite{majeed2021big}.

Azamfar et al. (2020) introduce a deep learning-based domain adaptation method designed for semiconductor manufacturing fault diagnosis. This method seeks to address variations in manufacturing processes that may affect data-driven approaches. Leveraging a deep neural network and the maximum mean discrepancy metric, the approach optimizes data representation effectively. Experimental results on a real-world semiconductor manufacturing dataset showcase the method's effectiveness and generalization in quality inspection. However, the paper needs to delve into the potential limitations of this proposed method~\cite{azamfar2020deep}.

Qu et al. (2019) offer a comprehensive overview of intelligent manufacturing systems (SMSs), which have garnered substantial attention from countries and manufacturing enterprises. The paper encompasses the evolution, definition, objectives, functional requirements, business requirements, technical requirements, and components of SMSs. Additionally, it presents a model for autonomous SMSs driven by dynamic demand and critical performance indicators. However, the paper does not discuss the challenges, limitations, or potential risks related to data privacy and security that may arise during the implementation of SMSs. Nevertheless, the findings contribute to a thorough understanding of 
SMSs and serve as a valuable reference for manufacturing enterprises looking to transition to intelligent systems~\cite{qu2019smart}.
\textbf{Table}~\ref{tab:comparison} summarizes some of the existing literature that employs Machine Learning (ML) and Deep Learning (DL) based approaches in the context of additive manufacturing (AM).
\section{Motivation}\label{mot}
Localizing specific regions in computer vision is relatively challenging compared to classifying objects~\cite{bappy2017exploiting, zou2023object}. 
\begin{table*}[]
\caption{Comparison of Existing Literature Reviews on ML/DL Approaches}
\label{tab:comparison}
\centering
\resizebox{.9\textwidth}{!}{%
\begin{tabular}{@{}lllcll@{}}
\toprule
\textbf{Reference} & \textbf{Contributions} & \textbf{ML/DL Methods} & \textbf{\begin{tabular}[c]{@{}c@{}}Explainable\\ AI\end{tabular}} & \textbf{Advantages} & \textbf{Limitations} \\ \midrule
\textbf{\begin{tabular}[c]{@{}l@{}} \cite{azamfar2020deep} \end{tabular}} & \begin{tabular}[c]{@{}l@{}}Domain adaptation for\\ semiconductor \\ fault diagnosis\end{tabular} & \begin{tabular}[c]{@{}l@{}}Deep Neural \\ Network (DNN)\end{tabular} & No & \begin{tabular}[c]{@{}l@{}}Effective quality \\ inspection\end{tabular} & \begin{tabular}[c]{@{}l@{}}Limited discussion \\ on method \\ limitations\end{tabular} \\
\textbf{\begin{tabular}[c]{@{}l@{}} \cite{qu2019smart} \end{tabular}} & \begin{tabular}[c]{@{}l@{}}Overview of smart \\ manufacturing systems\end{tabular} & \begin{tabular}[c]{@{}l@{}}Machine \\ Learning (ML)\end{tabular} & No & \begin{tabular}[c]{@{}l@{}}Comprehensive \\ understanding\end{tabular} & \begin{tabular}[c]{@{}l@{}}Lacks discussion \\ on implementation \\ challenges\end{tabular} \\
\textbf{\begin{tabular}[c]{@{}l@{}} \cite{wang2020machine} \end{tabular}} & \begin{tabular}[c]{@{}l@{}}Review of ML \\ applications \\ in additive \\ manufacturing\end{tabular} & ML & No & \begin{tabular}[c]{@{}l@{}}Insights into AM \\ potential\end{tabular} & \begin{tabular}[c]{@{}l@{}}No discussion of \\ ML limitations\end{tabular} \\
\textbf{\begin{tabular}[c]{@{}l@{}} \cite{qi2019applying} \end{tabular}} & \begin{tabular}[c]{@{}l@{}}NNs for complex \\ pattern recognition \\ in AM\end{tabular} & \begin{tabular}[c]{@{}l@{}}Neural \\ Networks (NNs)\end{tabular} & No & \begin{tabular}[c]{@{}l@{}}Potential for agile \\ manufacturing\end{tabular} & \begin{tabular}[c]{@{}l@{}}Limited specificity \\ on defects \\ predicted\end{tabular} \\
\textbf{\begin{tabular}[c]{@{}l@{}} \cite{kwon2020deep} \end{tabular}} & \begin{tabular}[c]{@{}l@{}}Deep learning for \\ melt-pool image \\ classification\end{tabular} & \begin{tabular}[c]{@{}l@{}}Deep Neural \\ Network\end{tabular} & No & \begin{tabular}[c]{@{}l@{}}Effective for \\ monitoring\end{tabular} & \begin{tabular}[c]{@{}l@{}}No discussion of \\ DL limitations\end{tabular} \\
\textbf{\begin{tabular}[c]{@{}l@{}} \cite{aquil2020predicting} \end{tabular}} & \begin{tabular}[c]{@{}l@{}}Software defect \\ prediction \\ using ML \\ techniques\end{tabular} & \begin{tabular}[c]{@{}l@{}}Various ML \\ techniques\end{tabular} & No & \begin{tabular}[c]{@{}l@{}}Consistent high-\\ accuracy \\ predictions\end{tabular} & \begin{tabular}[c]{@{}l@{}}Limited defect \\ type specificity\end{tabular} \\
\textbf{\begin{tabular}[c]{@{}l@{}} \cite{dogan2021systematic} \end{tabular}} & \begin{tabular}[c]{@{}l@{}}Literature review \\ on ML in \\ manufacturing\end{tabular} & \begin{tabular}[c]{@{}l@{}}ML, \\ Data \\ Mining\end{tabular} & No & \begin{tabular}[c]{@{}l@{}}Highlights \\ ML benefits\end{tabular} & \begin{tabular}[c]{@{}l@{}}Lacks detailed \\ analysis of \\ limitations\end{tabular} \\
\textbf{\begin{tabular}[c]{@{}l@{}} \cite{majeed2021big} \end{tabular}} & \begin{tabular}[c]{@{}l@{}}BD-SSAM \\ framework for \\ sustainable AM\end{tabular} & DNN & No & \begin{tabular}[c]{@{}l@{}}Effective \\ energy control\end{tabular} & \begin{tabular}[c]{@{}l@{}}Focus on early \\ lifecycle \\ and specific \\ material\end{tabular} \\ \bottomrule
\end{tabular}%
}
\end{table*}
While Convolutional Neural Network (CNN) based approaches can easily classify between two classes, identifying specific regions can be difficult for the same CNN model~\cite{wahab2017two,xue2018fast}.In early study~\cite{ahsan2023defect}, it was observed that the proposed Transfer Learning (TL)-based approach could accurately classify defect and non-defect 3D printed cylinder images. 

\begin{figure}[ht]
    \centering
    \includegraphics[width=.6\textwidth]{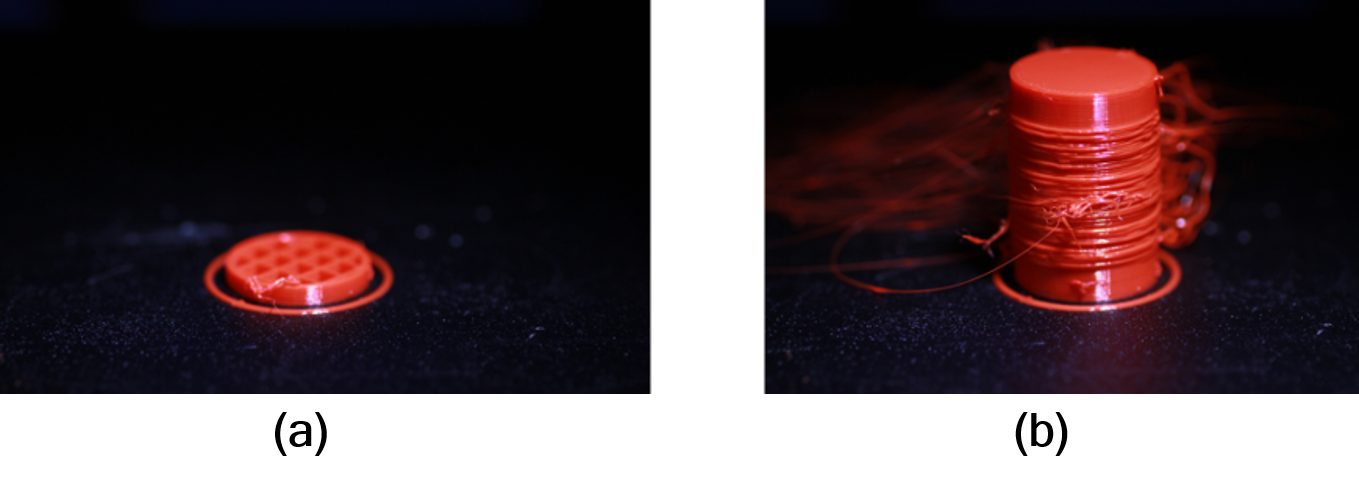}
    \caption{Sample cylinder image with defect (a) early stage and (b) final stage of 3D printing.}
    \label{fig:3dsam}
\end{figure}

However, these models often failed to identify the defect region of the cylinder images. The dataset consisted of images captured at different stages of the 3D printing process, with each image having varying sizes, regions, and pixel densities, as illustrated in \textbf{Figure}~\ref{fig:3dsam}. \textbf{Figure}~\ref{fig:3dsam}(a) shows the cylinder image at the early stage of printing, while \textbf{Figure}~\ref{fig:3dsam}(b) displays the complete cylinder image. It can be observed that for the TL model in \textbf{Figure}~\ref{fig:3dsam}(a), there are many dark and unnecessary regions, while in \textbf{Figure}~\ref{fig:3dsam}(b), the entire cylinder image is present, which allows the model to learn more in detail. As a result, such discrepancies and unnecessary regions may cause the proposed TL-based models to be unable to identify the defect regions. This situation is distinct from scenarios involving static images, where both the size and region remain consistent. Such consistency poses additional challenges when attempting to identify defect regions within images of 3D-printed cylinders, unlike the uniformity observed in images like standard medical or natural images, which maintain a consistent size~\cite{ahsan2024enhancing, nayem2023few, ahsan2023invariant}.

Finally, the interpretation of model predictions is crucial to understanding the behavior of the proposed models. Local Interpretable Model-Agnostic Explanations (LIME) and Gradient-weighted Class Activation Mapping (Grad-CAM) approaches will be applied to accomplish this. These techniques enable identifying essential features in an input image that contributed to the model's prediction. By generating heatmaps, researchers can gain insight into how the model arrived at its decision, which can provide an understanding of the model's inner workings.

\section{Proposed Approahces}\label{proa}
\begin{figure*}[ht]
    \centering
    \includegraphics[width=.8\textwidth]{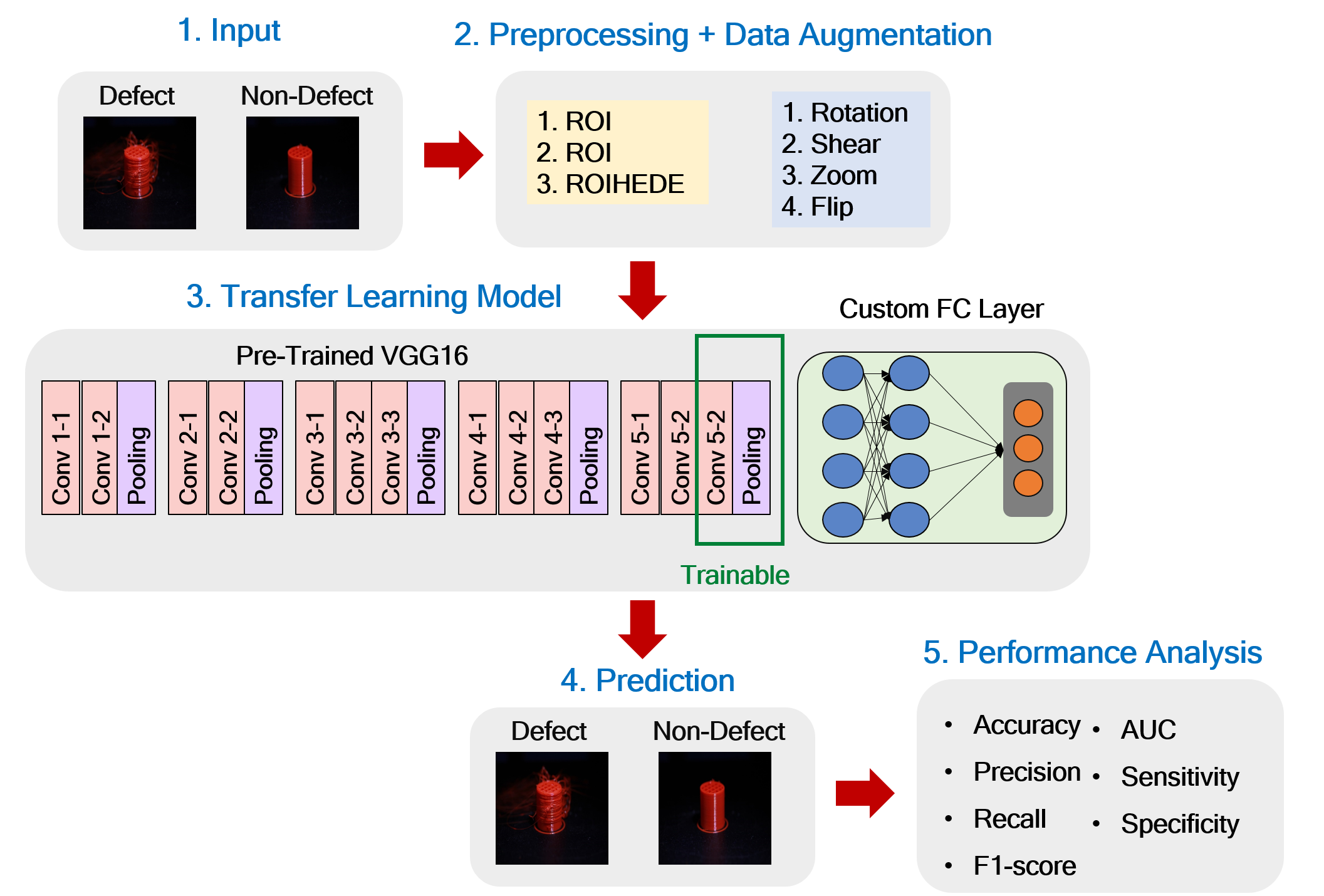}
    \caption{Flow diagram of the proposed models, including several steps: 1. \textbf{Input}—defect and non-defect samples, 2. \textbf{Preprocessing} (ROI, HE, DE) and \textbf{Data augmentation} (Rotation, Shear, Zoom, Flip, etc.), 3. \textbf{Transfer learning model}—VGG16 with a custom fully connected (FC) layer, 4. \textbf{Prediction}—classified as defect or non-defect, 5. \textbf{Performance analysis} with various statistical measurements.}
    \label{fig:propp}
\end{figure*}

The proposed approach in this study involves image preprocessing, data preparation, the use of updated transfer learning models with custom layers, and details about the model training procedure (as shown in \textbf{Figure}~\ref{fig:propp}), which are discussed below.
\subsection{Data Collection}~\label{data}
\begin{figure}[ht]
\centering
    \includegraphics[width=.7\textwidth]{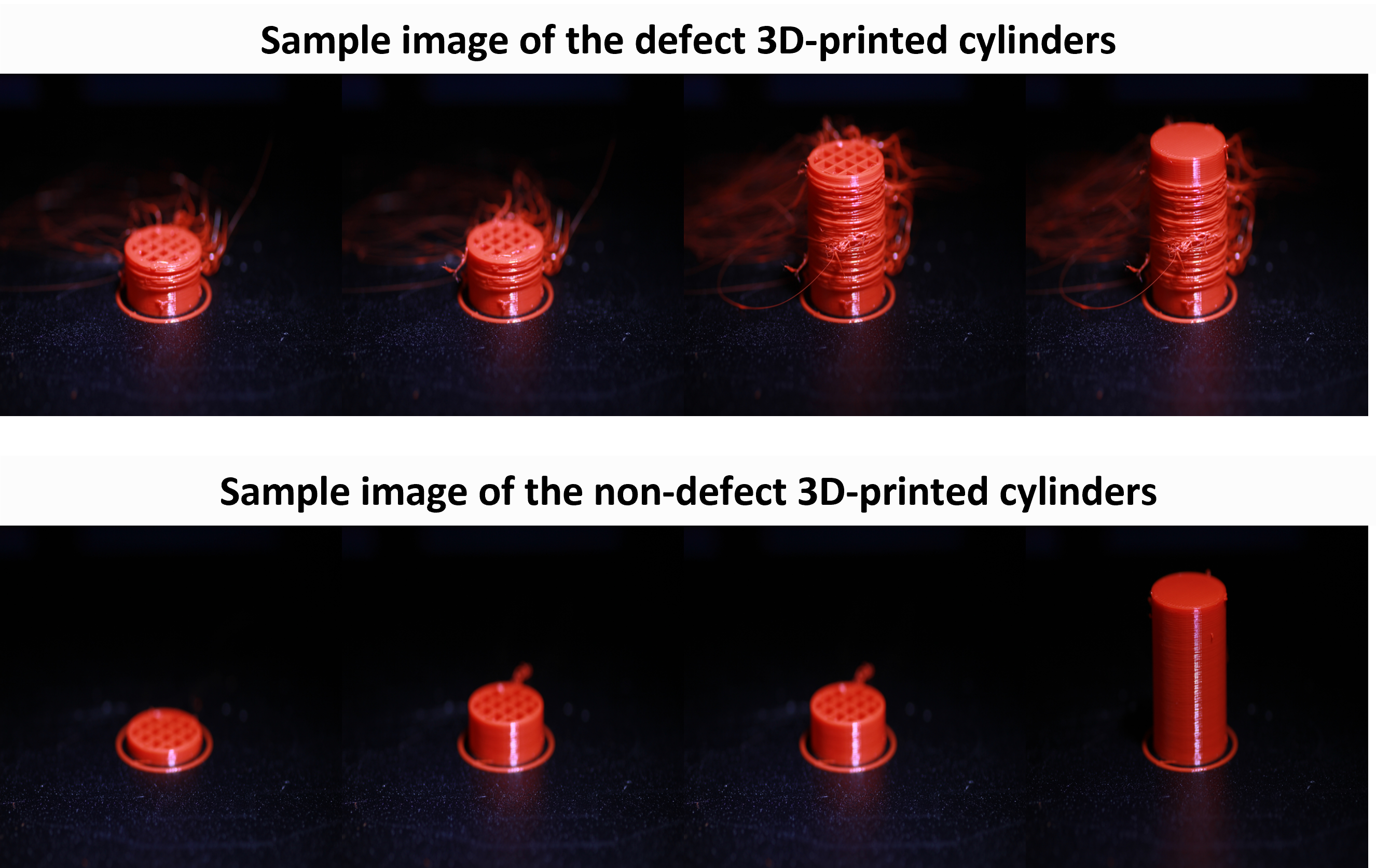}
    \caption{Representative samples of 3D-printed product images obtained from SMIS, which have been utilized in this research.}
    \label{fig:3ds}
\end{figure}

The dataset was generously provided by the Smart Materials and Intelligent Systems (SMIS) Laboratory. It comprises a collection of 3D printed prototype images of cylindrical objects, some of which contain defects while others are non-defective.
\textbf{Figure}~\ref{fig:3ds} displays a selection of typical images of cylindrical objects from the dataset, including both defective and non-defective examples.

\textbf{Table}~\ref{tab:3d} provides an overview of the allocation of data for training and testing of each of the CNN models that were investigated. In both studies, six different DL approaches were investigated: VGG16~\cite{simonyan2014very}, InceptionResNetV2~\cite{szegedy2017inception}, ResNet50~\cite{akiba2017extremely}, MobileNetV2~\cite{sandler2018mobilenetv2}, ResNet101~\cite{he2016deep} and VGG19~\cite{simonyan2014very}.

\begin{table}[h]
\centering
\caption{Allocation of data for training and testing of TL models used in this study.}
\label{tab:3d}
\resizebox{.45\textwidth}{!}{%
\begin{tabular}{@{}lllll@{}}
\toprule
\textbf{Dataset} & \textbf{Label} & \textbf{Train} & \textbf{Test} & \textbf{Total} \\ \midrule
\multirow{2}{*}{Study One} & Defect & 105 & 31 & 136 \\
 & Non-defect & 112 & 24 & 136 \\
\multirow{2}{*}{Study Two} & Defect & 736 & 211 & 947 \\
 & Non-defect & 1677 & 393 & 2070  \\ \bottomrule
\end{tabular}%
}
\end{table} 
\subsection{Image Pre-Processing}
In our proposed approach, we used three image preprocessing techniques: Region of Interest (ROI), Histogram Equalization (HE), and Details Enhancer (DE).
\subsubsection{ROI}
ROI selection is used to identify and extract the region of interest from the input image. This technique involves selecting an image's specific area or region to analyze or modify, allowing for more efficient image processing. The ROI is determined by calculating the mean intensity value of the image and selecting a specific threshold value for the selected region. The mathematical formula for ROI selection can be represented as~\cite{erdem_2020, girshick2015fast}:

\begin{equation}
I_{ROI} = \begin{cases}
I(x,y), & \text{if } I(x,y) > T \\
0, & \text{otherwise}
\end{cases}
\end{equation}

where $I_{ROI}$ is the ROI selected from the input image $I$, $(x,y)$ represents the pixel coordinates of the image, and $T$ is the threshold value for the selected region.
\subsubsection{HE}
The second pre-processing step, HE, is a widely-used image processing technique that enhances images' contrast by redistributing pixel values. This technique improves the visual quality of images by normalizing the intensity levels, which improves image detail and clarity. The mathematical formula for HE can be represented as~\cite{joshi_2022, dadhich2018practical}:
\begin{equation}
g(i,j) = \frac{CDF(f(i,j)) - CDF_{min}}{(MN - CDF_{min})} \times (L-1)
\end{equation}

where $g(i,j)$ is the output image after HE, $f(i,j)$ is the input image, $CDF$ is the cumulative distribution function, $CDF_{min}$ is the minimum cumulative distribution function value, $MN$ is the total number of pixels in the image, and $L$ is the maximum pixel intensity value.
\begin{figure}[ht]
    \centering
    \includegraphics[width=.7\textwidth]{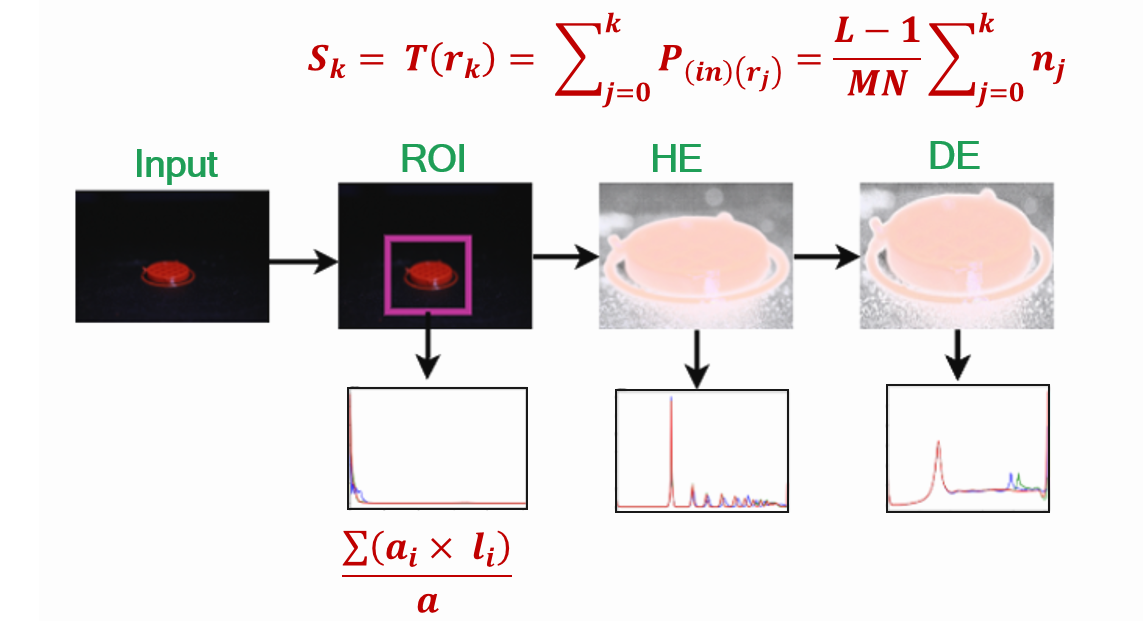}
    \caption{Preprocessing steps to identify defect regions before training the proposed CNN model.}
    \label{fig:pre}
\end{figure}
\subsubsection{DE}
The third pre-processing step, DE, is a technique used to enhance the details of an image by increasing its contrast and sharpness. This technique is particularly useful when dealing with low-contrast images, as it can help to bring out more subtle details that might otherwise be difficult to see. The mathematical formula for DE can be represented as~\cite{dadhich2018practical, bansal_2022}:
\begin{equation}
g(i,j) = (1 + k \cdot (f(i,j) - 128)) \cdot f(i,j)
\end{equation}

where $g(i,j)$ is the output image after DE, $f(i,j)$ is the input image, and $k$ is a constant value that determines the degree of enhancement.

These pre-processing steps are then combined with the proposed modified VGG16 CNN architecture and tested on a designated dataset. These pre-processing techniques are expected to improve the accuracy and efficiency of the proposed model by enhancing the input image data and enabling more accurate defect region identification in 3D-printed cylinder images. 

The pre-processing steps are integrated with the proposed CNN models as illustrated in \textbf{Figure}~\ref{fig:pre}. In the figure, First formula The first formula calculates the cumulative distribution function of an input image denoted as $S_k$. It uses a transformation function T to map pixel intensities from the input image to the output image. The formula uses the k-th gray level, $r_k$, and the normalized histogram of the input image, $P_in(r_j)$, where L represents the total number of possible intensity levels, M represents the number of rows of the image, and N represents the number of columns of the image. It calculates the mapping function for histogram equalization, which redistributes the pixel intensity values in an image to produce a uniform histogram. The second formula calculates the average length of a set of line segments. It uses the length of the i-th line segment, $a_i$, the number of occurrences of the j-th length in the set of line segments, $l_j$, and the total number of line segments, A. The formula multiplies each line segment length, $a_i$, by the number of occurrences, $l_j$, and adds up these products for all lengths, j. This gives us the sum of the products of the length and frequency of each line segment, which is then divided by the total number of line segments, A, to give the average length of the set of line segments.

\subsection{Data Preparation}
In the data preparation phase, raw data is transformed into a format more suitable for creating an effective model. Data preparation involves key processes including data normalization, feature selection, and data augmentation, all of which are essential for finalizing the dataset for training.
\subsection{Data Normalization}
Data normalization is a crucial preprocessing step in our study, aiming to scale the features of the dataset to a common range without distorting differences in the ranges of values or losing information. This standardization enhances the performance and convergence speed of the machine learning algorithms. In our research, we have employed the StandardScaler technique, which standardizes features by removing the mean and scaling to unit variance. This approach adjusts the distribution of each attribute to have a mean of zero and a standard deviation of one, following the formula:

\begin{equation}
z = \frac{x - \mu}{\sigma}
\end{equation}

where \(z\) represents the standardized value, \(x\) is the original value, \(\mu\) is the mean of the feature values, and \(\sigma\) is the standard deviation of the feature values. The application of the StandardScaler ensures that our model handles the input variables uniformly, thereby improving the modeling outcomes.
\subsection{Feature Extraction}
Feature extraction plays a pivotal role in enhancing the performance of convolutional neural networks (CNNs) for image classification tasks. In our study, we employed feature extraction techniques to distill relevant information from images, thereby simplifying the input data for our CNN models. This process reduces computational complexity and improves model efficiency by focusing on essential attributes that contribute to the accuracy of the classification. Specifically, we leveraged these techniques to facilitate the development of transfer learning models. Transfer learning allows our models to utilize knowledge acquired from one task and apply it to a different but related task, significantly enhancing learning efficiency and model performance with limited data. A fundamental equation in feature extraction for CNNs, particularly in the context of transfer learning, is the formulation of feature vectors:

\begin{equation}
F = CNN_{\theta}(I)
\end{equation}

where \(F\) represents the extracted feature vector, \(CNN_{\theta}\) denotes the pre-trained CNN model with parameters \(\theta\), and \(I\) is the input image. By applying this technique, our study efficiently harnesses pre-trained models to achieve superior image classification results, showcasing the effectiveness of feature extraction in streamlining the training process and boosting the predictive capabilities of transfer learning models.
\subsection{Data Augmenation}
Data augmentation plays a pivotal role in enhancing the size and variability of our training dataset, crucial for improving the performance of convolutional neural network (CNN) models, particularly in the development of transfer learning models for image classification. By artificially expanding the dataset through various transformations, we can simulate a wider range of scenarios that the model might encounter in the real world, thus improving its robustness and accuracy.

In our study, we utilized the `ImageDataGenerator` from Keras to implement data augmentation, which dynamically applies a series of transformations to the input images during the training phase. This method not only enriches our dataset but also introduces a form of regularization, helping to prevent overfitting by providing a more diverse set of training examples. \textbf{Table}~\ref{tab:data_augmentation} summarizes the different data augmentation approaches we have adopted, reflecting the parameters and actions defined in our `ImageDataGenerator` configuration.
\begin{table*}[]
\caption{Overview of data augmentation approaches used in this study.}
\label{tab:data_augmentation}
\centering
\resizebox{.7\textwidth}{!}{%
\begin{tabular}{@{}lll@{}}
\toprule
\textbf{Augmentation} & \textbf{Value} & \textbf{Approaches} \\ \midrule
Rotation Range & 15 & Input data is rotated between $-15^\circ$ and $15^\circ$ \\
Height Shift Range & 0.1 & Vertically shifts the image by 10\% \\
Width Shift Range & 0.1 & Horizontally shifts the image by 10\% \\
Shear Range & 0.2 & Applies a shearing transformation of 0.2 \\
Zoom Range & 0.2 & Zooms in or out by 20\% from the center \\
Horizontal Flip & True & Randomly flips the image horizontally \\
Vertical Flip & True & Randomly flips the image vertically \\
Fill Mode & "nearest" & Fills in new pixels with the nearest pixel values \\ \bottomrule
\end{tabular}%
}
\end{table*}
\subsection{Proposed Transfer Learning Approaches}
In this study, we have implemented transfer learning techniques by utilizing a modified VGG16 model as our foundation. Transfer learning allows us to leverage pre-trained models, which have been previously trained on a large dataset (e.g., ImageNet), to enhance the performance on our specific task without the need for training from scratch. This approach significantly reduces the computation time and resources required, while also potentially improving accuracy due to the pre-learned features.

\subsubsection{Base Model Configuration}

The foundation of our model leverages the VGG16 architecture, a widely recognized convolutional neural network model proposed by Simonyan and Zisserman. The VGG16 model, known for its simplicity and depth, has shown remarkable performance in large-scale image recognition tasks. 
In our adaptation, we configure VGG16 without its original classification layers (commonly referred to as "top layers") to serve as a feature extractor for inputs of size \(224 \times 224 \times 3\), which represents height, width, and color channels of the input images respectively. This can be mathematically represented as:

\begin{equation}
    \text{Input Layer Size} = 224 \times 224 \times 3
\end{equation}

\subsubsection{Custom Model Layers}

\begin{figure}[ht]
    \centering
    \includegraphics[width=.4\textwidth]{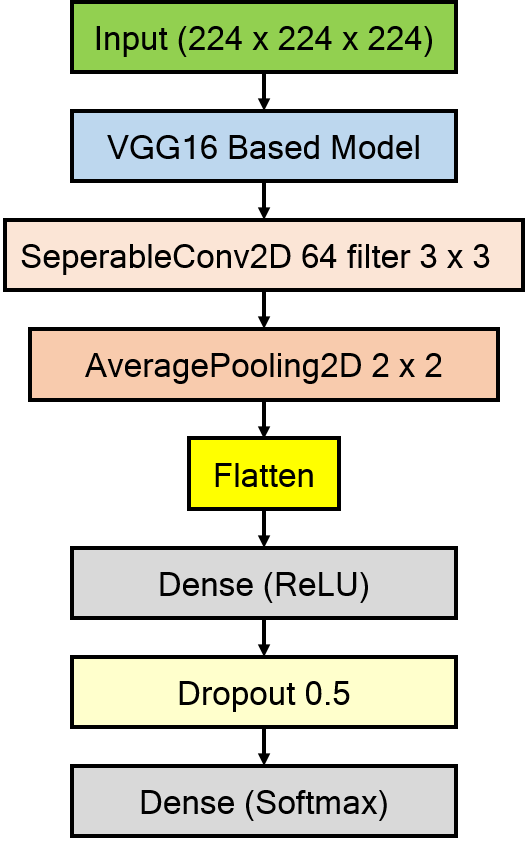}
    \caption{Flow diagram of the proposed transfer learning models, including the base model \textbf{VGG16}, followed by a \textbf{separable convolutional} layer, \textbf{average pooling}, and \textbf{flattening}, culminating in two \textbf{dense} layers with a dropout rate of \underline{\textbf{0.5}}.}
    \label{fig:tfff}
\end{figure}

After the base VGG16 model processes the input, the model's output serves as the input to our customized layers tailored for our specific task. The modifications include:

\begin{enumerate}

    \item \textbf{Separable Convolution Layer (SeparableConv2D)}: This layer performs a depthwise spatial convolution followed by a pointwise convolution, reducing the number of parameters compared to a standard convolution layer. It is defined by 64 filters of size \(3 \times 3\). The mathematical operation can be represented as:

    \begin{equation}
        \text{SeparableConv2D}: F_{out} = F_{dw}(I) * F_{pw}(K)
    \end{equation}
    
    where \(F_{out}\) is the output feature map, \(F_{dw}(I)\) is the depthwise convolution on input \(I\), and \(F_{pw}(K)\) is the pointwise convolution applying \(K\) filters.
    
    \item \textbf{Average Pooling Layer (AveragePooling2D)}: This layer reduces the spatial dimensions (height and width) of the input feature map by performing average pooling, using a pool size of \(2 \times 2\). The operation can be described as:
    
    \begin{equation}
        \text{Average Pooling}: P_{avg}(F_{out}) = \frac{1}{N} \sum_{i=1}^{N} x_i
    \end{equation}
    
    where \(P_{avg}\) is the average pooled output, \(F_{out}\) is the input feature map, \(N\) is the number of elements in the pooling window, and \(x_i\) are the elements within the window.
    
    \item \textbf{Flatten Layer}: This layer converts the pooled feature map into a single vector, facilitating a transition from convolutional layers to fully connected layers. Mathematically, it reshapes the input tensor \(T\) of shape \(a \times b \times c\) into a vector \(V\) of size \(n\), where \(n = a \cdot b \cdot c\).
    
    \item \textbf{Fully Connected Layer (Dense)}: A dense layer with 64 units follows, applying a ReLU activation function to introduce non-linearity. It is represented as:
    
    \begin{equation}
        \text{Dense Layer Output} = \text{ReLU}(W \cdot V + b)
    \end{equation}
    
    where \(W\) and \(b\) are the weights and biases of the layer, and \(V\) is the input vector from the Flatten layer.
    
    \item \textbf{Dropout Layer}: To prevent overfitting, a dropout layer with a rate of 0.5 is employed, randomly setting a fraction of input units to 0 at each update during training time.
    
    \begin{equation}
        \text{Dropout}: D(V, r)
    \end{equation}
    
    where \(D\) is the dropout function, \(V\) is the input vector, and \(r\) is the dropout rate.
    
    \item \textbf{Output Layer (Dense)}: The final layer is a densely connected layer with 2 units and a softmax activation function, designed for binary classification tasks. The softmax function converts the output scores into probabilities:
    
    \begin{equation}
        \text{Softmax}(z_i) = \frac{e^{z_i}}{\sum_{j} e^{z_j}}
    \end{equation}
    
    where \(z_i\) is the input to the softmax function for class \(i\), and the denominator is the sum of exponential scores for all classes. 
\end{enumerate}
\textbf{Figure}~\ref{fig:tfff} presents a schematic diagram of our custom models and their corresponding layers.
\subsubsection{Optimizer}
In this study, we employed the Adam optimizer for model optimization. Adam, an algorithm for first-order gradient-based optimization of stochastic objective functions, stands out for its efficiency with large problems involving a lot of data or parameters. It combines the advantages of two other extensions of stochastic gradient descent, specifically Adaptive Gradient Algorithm (AdaGrad) and Root Mean Square Propagation (RMSProp). Adam calculates an exponential moving average of the gradient and the squared gradient, and the parameters beta1 and beta2 control these moving averages.

The equation governing the update rule for the Adam optimizer is:

\begin{equation}
\theta_{t+1} = \theta_t - \frac{\eta}{\sqrt{\hat{v}_t} + \epsilon} \hat{m}_t
\end{equation}

where:
- \(\theta\) represents the parameters,
- \(\eta\) is the learning rate,
- \(\hat{m}_t\) and \(\hat{v}_t\) are estimates of the first moment (the mean) and the second moment (the uncentered variance) of the gradients, respectively,
- \(\epsilon\) is a small scalar added to improve numerical stability.
\subsubsection{Loss Function}
For the loss function, we selected binary crossentropy. This choice is appropriate for binary classification problems, where the goal is to assign each input sample to one of two classes. Binary crossentropy measures the distance between the probability distributions of the predicted and actual values, with the aim of minimizing this distance during training. The loss function is defined as:

\begin{equation}
L(y, \hat{y}) = -\frac{1}{N} \sum_{i=1}^{N} [y_i \log(\hat{y}_i) + (1 - y_i) \log(1 - \hat{y}_i)]
\end{equation}

where:
- \(L\) is the loss,
- \(N\) is the number of samples,
- \(y_i\) is the actual label of sample \(i\),
- \(\hat{y}_i\) is the predicted probability of sample \(i\) belonging to the positive class.

\subsection{Model Training}
The entire experiment was conducted using an office-grade laptop equipped with standard specifications, which encompassed a Windows 10 operating system, an Intel Core i7-7500U processor, and 16 GB of RAM. During this study, we observed training accuracy and loss throughout the epochs. We trained the model and monitored its performance closely. We used data augmentation to enhance the model's generalization from training to unseen data, which also increased its robustness. The addition of variability through augmented data helped prevent overfitting. This approach to training, along with data augmentation, was fundamental to our methodology.
\subsection{Computational Complexity}
We assessed computational complexity by calculating the process time, marking the start (\( t_1 \)) and end (\( t_2 \)) of the training session. The difference, \( T = t_2 - t_1 \), indicates the total training time, providing a measure of the computational resources used. This metric is vital for evaluating the training efficiency.
\subsection{Optimal Parameters}
Three parameters of the deep learning model are optimized in this study: batch size (which determines the number of samples processed before updating internal model parameters), epochs (indicating the number of times the learning algorithm iterates over the entire dataset), and learning rate (a hyperparameter controlling the magnitude of model adjustments when updating model weights to calculate error)~\cite{brownlee2018difference}.

Inspired by prior works~\cite{smith2018disciplined,smith2017don}, a grid search method commonly employed for parameter tuning was applied in this study~\cite{bergstra2012random}. Initially, the following parameters were randomly selected:
\begin{center}
Batch\, size = $[4,5,8,10]$\\
Number\, of\, epochs = $[10,20,30,40]$\\
Learning\, rate = $[.001, .01, 0.1]$
\end{center}

For Study One, better results were achieved using the grid search method with the following parameters:

\begin{center}
Batch\, size = $8$\\
Number\, of\, epochs = $30$\\
Learning\, rate = $.001$\\
\end{center}
Similarly, for Study Two, the best results were achieved with:
\begin{center}
Batch\, size = $50$\\
Number\, of\, epochs = $50$\\
Learning\, rate = $.001$\\
\end{center}

The optimization algorithm employed for all models was Adam, an adaptive learning rate optimization algorithm recognized for its robust performance in binary image classification\cite{perez2017effectiveness,filipczuk2013computer}. Adhering to the customary procedure in data mining techniques, 80\% of the accessible data was designated for training, while the remaining 20\% was set aside for testing purposes~\cite{mohanty2016using, menzies2006data, stolfo2000cost}.
\subsection{Performance Assessment}
The assessment of performance was conducted utilizing diverse metrics, encompassing model accuracy, precision, recall, and F1-score~\cite{ahsan2020face}. 

\begin{equation}
   \textrm{Accuracy} =\frac{t_p + t_n}{t_p + t_n + f_p + f_n} 
\end{equation}

\begin{equation}
 \textrm{Precision} = \frac{t_p}{t_p+f_p}   
\end{equation}

\begin{equation}
  \textrm{Recall (Sensitivity)}=\frac{t_p}{t_p + f_n}  
\end{equation}

\begin{equation}
    \textrm{F1-score} =2\times\frac{\textrm{Precision}\times\textrm{Recall}}{\textrm{Precision+Recall}}
\end{equation}

\begin{equation}
    \textrm{Specificity} = \frac{t_n}{t_n + f_p}
\end{equation}

Where,
\begin{itemize}
\item True Positive ($t_p$)= Defect cylinder classified as Defect
\item False Positive ($f_p$)= Normal cylinder classified as Defect
\item True Negative ($t_n$)= Normal cylinder classified as Normal
\item False Negative ($f_n$)= Defect cylinder classified as Normal.
\end{itemize}
\section{Results}\label{observation}
\subsection{Study One}
\textbf{Table}~\ref{tab:std1} shows the performance of different proposed CNN models, including Modified VGG16, ROI Selection (ROIN), ROI Selection with Histogram Equalization (ROIHEN), and ROI Selection with Histogram Equalization and Details Enhancer (ROIHEDEN) on the train and test set of the 3D printed cylinder image small dataset used in this study. The results show that all the proposed approaches achieve promising results, with accuracies ranging from 0.94 to 1 on the test set. Additionally, all the models achieved perfect precision, recall, and F1-score values, indicating no false positives or negatives in identifying the image's defect region. Furthermore, the specificity of all the models was also high, with values ranging from 0.8750 to 1, indicating a low false-positive rate.
\begin{table}[ht]
\centering
\caption{Performance of different CNN models on the training and test datasets from Study One with 95\% confidence interval.}
\label{tab:std1}
\resizebox{\textwidth}{!}{%
\begin{tabular}{@{}lcccccccccc@{}}
\toprule
\multirow{2}{*}{\textbf{Model}} & \multicolumn{2}{c}{\textbf{Accuracy}} & \multicolumn{2}{c}{\textbf{Precision}} & \multicolumn{2}{c}{\textbf{Recall}} & \multicolumn{2}{c}{\textbf{F1-score}} & \multicolumn{2}{c}{\textbf{Specificity}} \\ \cmidrule(l){2-11} 
 & \textbf{Train} & \textbf{Test} & \textbf{Train} & \textbf{Test} & \textbf{Train} & \textbf{Test} & \textbf{Train} & \textbf{Test} & \textbf{Train} & \textbf{Test} \\ \midrule
Modified VGG16 & 0.94 $\pm$ 0.03 & 1.00 $\pm$ 0.00 & 0.94 $\pm$ 0.03 & 1.00 $\pm$ 0.00 & 0.94 $\pm$ 0.02 & 1.00 $\pm$ 0.00 & 0.94 $\pm$ 0.03 & 1.00 $\pm$ 0.00 & 0.88 $\pm$ 0.04 & 1.00 $\pm$ 0.00 \\
ROIN & 0.97 $\pm$ 0.02 & 1.00 $\pm$ 0.00 & 0.97 $\pm$ 0.02 & 1.00 $\pm$ 0.00 & 0.97 $\pm$ 0.02 & 1.00 $\pm$ 0.00 & 0.97 $\pm$ 0.02 & 1.00 $\pm$ 0.00 & 0.95 $\pm$ 0.03 & 1.00 $\pm$ 0.00 \\
ROIHEN & 0.96 $\pm$ 0.03 & 1.00 $\pm$ 0.00 & 0.96 $\pm$ 0.03 & 1.00 $\pm$ 0.00 & 0.96 $\pm$ 0.03 & 1.00 $\pm$ 0.00 & 0.96 $\pm$ 0.03 & 1.00 $\pm$ 0.00 & 0.92 $\pm$ 0.04 & 1.00 $\pm$ 0.00 \\
ROIHEDEN & 0.96 $\pm$ 0.03 & 1.00 $\pm$ 0.00 & 0.96 $\pm$ 0.03 & 1.00 $\pm$ 0.00 & 0.96 $\pm$ 0.03 & 1.00 $\pm$ 0.00 & 0.96 $\pm$ 0.03 & 1.00 $\pm$ 0.00 & 0.92 $\pm$ 0.04 & 1.00 $\pm$ 0.00 \\ \bottomrule
\end{tabular}%
}
\end{table}

\begin{figure}[ht]
    \centering
    \includegraphics[width=.7\textwidth]{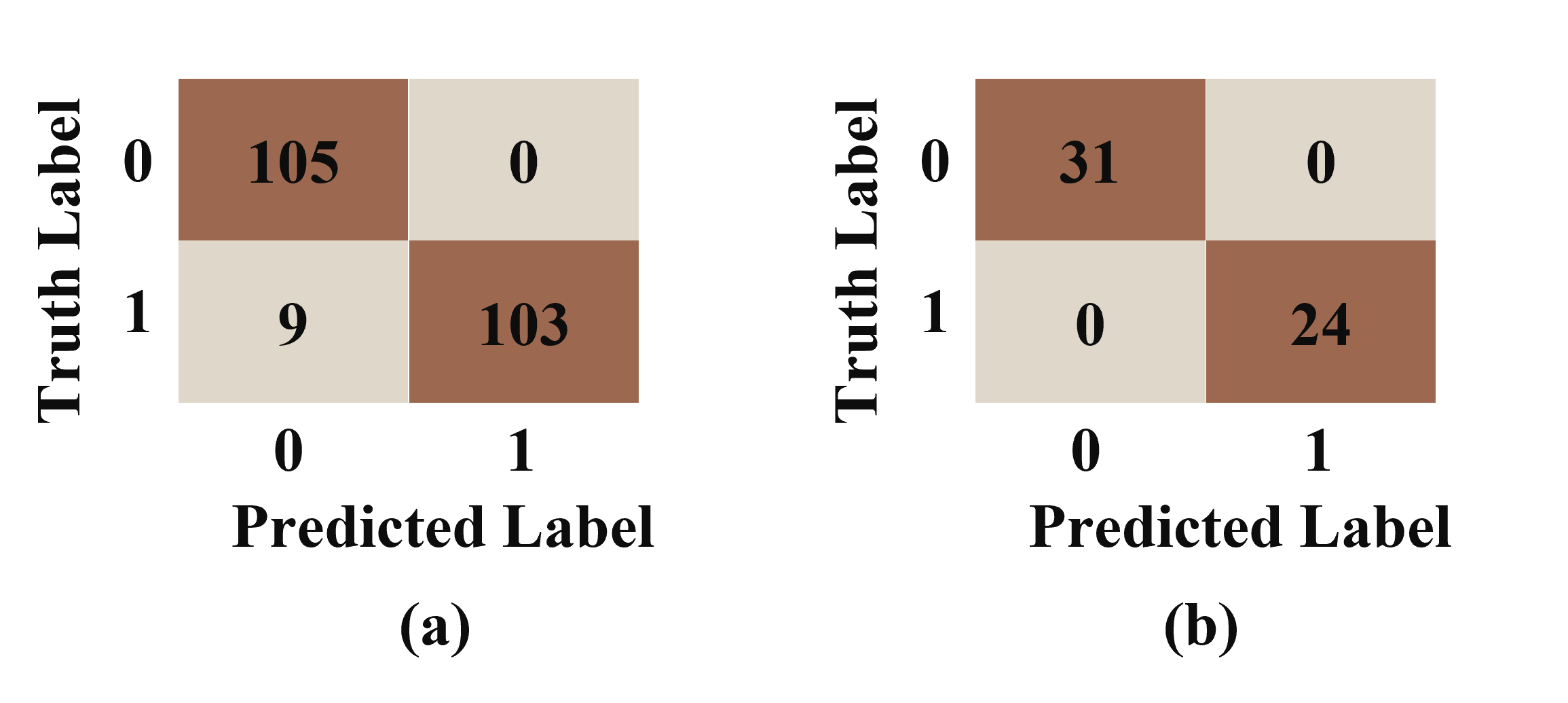}
    \caption{Confusion matrices of ROIHEDEN on (a) train and (b) test set.}
    \label{fig:con1}
\end{figure}

\textbf{Figure}~\ref{fig:con1} illustrates the confusion matrices of the proposed ROIHEDEN model on both the train and test sets. The figure shows that the proposed model misclassified only 9 samples from the train set (\textbf{Figure}~\ref{fig:con1}(a)) and zero samples from the test set (\textbf{Figure}~\ref{fig:con1}(b)).

\begin{figure}[ht]
    \centering
    \includegraphics[width=.7\textwidth]{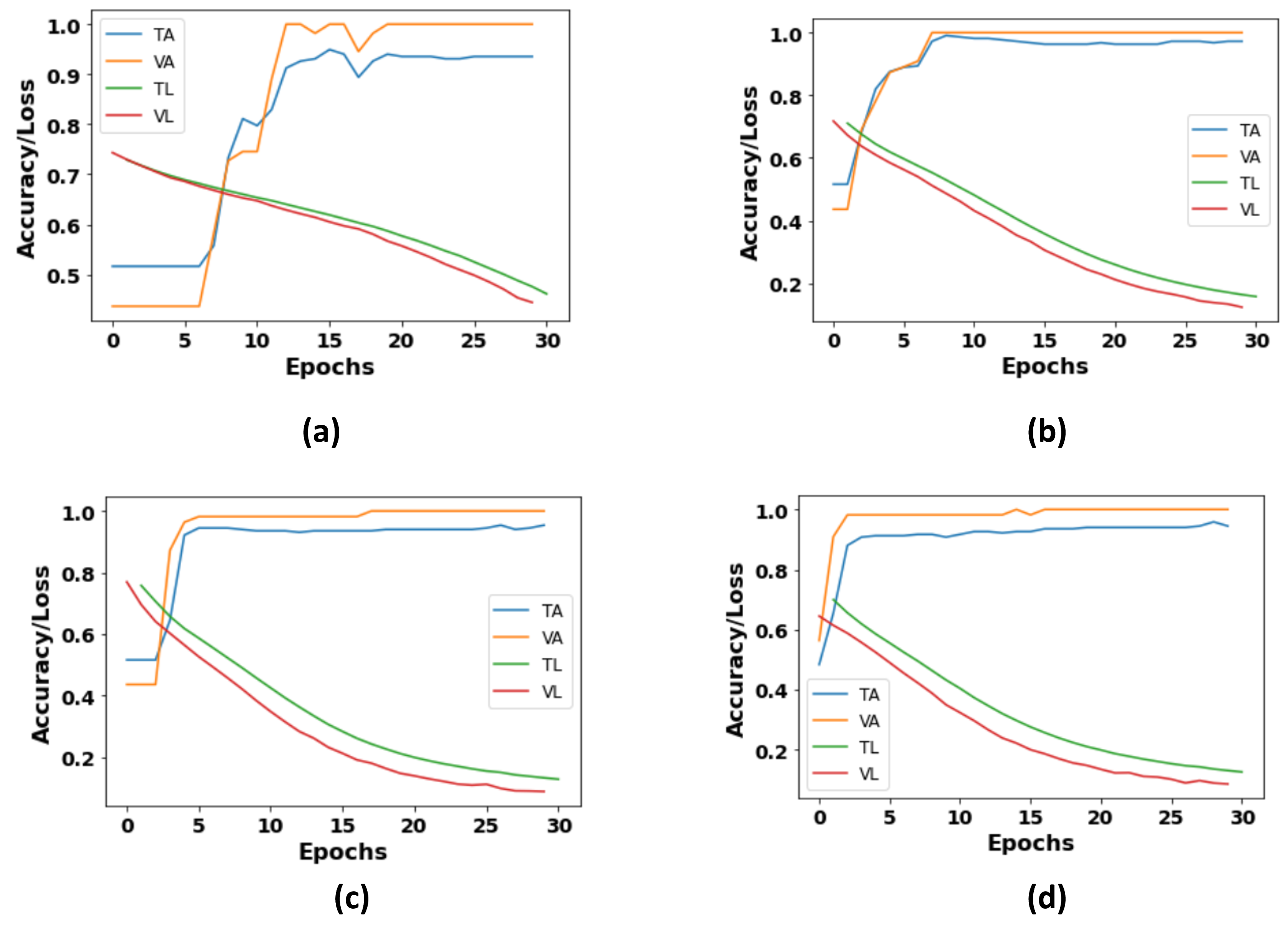}
    \caption{Training and validation performance throughout the training of (a) modified VGG16, (b) ROIN, (c) ROIHEN, and (d) ROIHEDEN during Study One. TA -- training accuracy; VA -- validation accuracy; TL -- training loss; VL -- validation loss.}
    \label{fig:tac1}
\end{figure}

In \textbf{Figure}~\ref{fig:tac1}, the performance of modified VGG16, ROIN, ROIHEN, and ROIHEDEN models during the training phase is compared. The results show that all of the proposed models continuously improved train accuracy and reduced train loss until epoch 30. 

\subsection{Study Two}
\textbf{Table}~\ref{tab:std2} presents the performance of the proposed CNN models on the train and test sets of the large dataset used in Study Two. The evaluation metrics used include accuracy, precision, recall, F1-score, and specificity. The models tested are the Modified VGG16, ROIN, ROIHEN, and ROHEDEN. From the table, it can be observed that the Modified VGG16 model achieved the highest accuracy and F1-score on both the train and test sets, while the ROIN model achieved the lowest accuracy and F1-score. However, the ROHEDEN model performed better than the Modified VGG16 model in terms of specificity on the test set. Overall, the results demonstrate that the proposed approaches of ROI selection, histogram equalization, and details enhancer can improve the performance of CNN models in detecting defects in 3D-printed cylinder images.

\begin{table*}[]
\centering
\caption{Proposed different CNN model's performance on the train and test set of the large dataset used during Study Two, with 95\% confidence interval.}
\label{tab:std2}
\resizebox{\textwidth}{!}{%
\begin{tabular}{@{}lllllllllll@{}}
\toprule
\multirow{2}{*}{\textbf{Model}} & \multicolumn{2}{c}{\textbf{Accuracy}} & \multicolumn{2}{c}{\textbf{Precision}} & \multicolumn{2}{c}{\textbf{Recall}} & \multicolumn{2}{c}{\textbf{F1-score}} & \multicolumn{2}{c}{\textbf{Specificity}} \\ \cmidrule(l){2-11} 
 & \textbf{Train} & \textbf{Test} & \textbf{Train} & \textbf{Test} & \textbf{Train} & \textbf{Test} & \textbf{Train} & \textbf{Test} & \textbf{Train} & \textbf{Test} \\ \midrule
Modified VGG16 & 0.97 $\pm$ 0.01 & 0.97 $\pm$ 0.01 & 0.97 $\pm$ 0.01 & 0.97 $\pm$ 0.01 & 0.97 $\pm$ 0.01 & 0.97 $\pm$ 0.01 & 0.96 $\pm$ 0.01 & 0.97 $\pm$ 0.01 & 0.89 $\pm$ 0.02 & 0.91 $\pm$ 0.02 \\
ROIN & 0.95 $\pm$ 0.01 & 0.96 $\pm$ 0.01 & 0.96 $\pm$ 0.01 & 0.96 $\pm$ 0.01 & 0.95 $\pm$ 0.01 & 0.96 $\pm$ 0.01 & 0.95 $\pm$ 0.01 & 0.96 $\pm$ 0.01 & 0.85 $\pm$ 0.03 & 0.88 $\pm$ 0.03 \\
ROIHEN & 0.94 $\pm$ 0.01 & 0.94 $\pm$ 0.01 & 0.95 $\pm$ 0.01 & 0.95 $\pm$ 0.01 & 0.94 $\pm$ 0.01 & 0.94 $\pm$ 0.01 & 0.94 $\pm$ 0.01 & 0.94 $\pm$ 0.01 & 0.81 $\pm$ 0.03 & 0.84 $\pm$ 0.03 \\
ROIHEDEN & 0.97 $\pm$ 0.01 & 0.96 $\pm$ 0.01 & 0.97 $\pm$ 0.01 & 0.96 $\pm$ 0.01 & 0.97 $\pm$ 0.01 & 0.96 $\pm$ 0.01 & 0.96 $\pm$ 0.01 & 0.96 $\pm$ 0.01 & 0.89 $\pm$ 0.02 & 0.90 $\pm$ 0.02 \\ \bottomrule
\end{tabular}%
}
\end{table*}

\begin{figure}[ht]
    \centering
    \includegraphics[width=.7\textwidth]{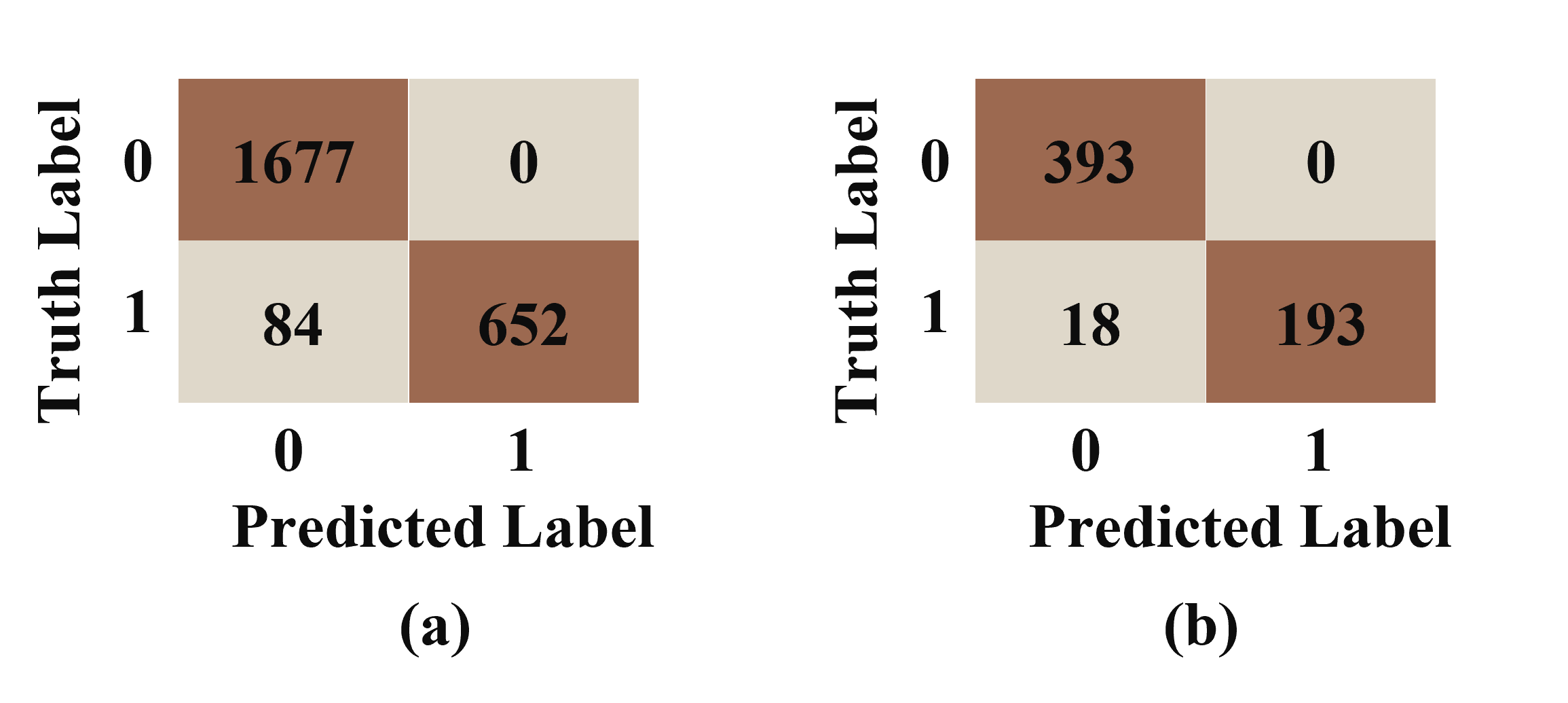}
    \caption{Confusion matrices of modified VGG16 on (a) train and (b) test set.}
    \label{fig:cons1}
\end{figure}

\textbf{Figure}~\ref{fig:cons1} illustrates the confusion matrices of the proposed modified VGG16 model on both the train and test sets during Study Two. The figure shows that the proposed model misclassified 11\% samples from the train set (\textbf{Figure}~\ref{fig:con1}(a)) and 8.5\% of the samples from the test set (\textbf{Figure}~\ref{fig:con1}(b)).

\begin{figure*}[ht]
    \centering
    \includegraphics[width=.7\textwidth]{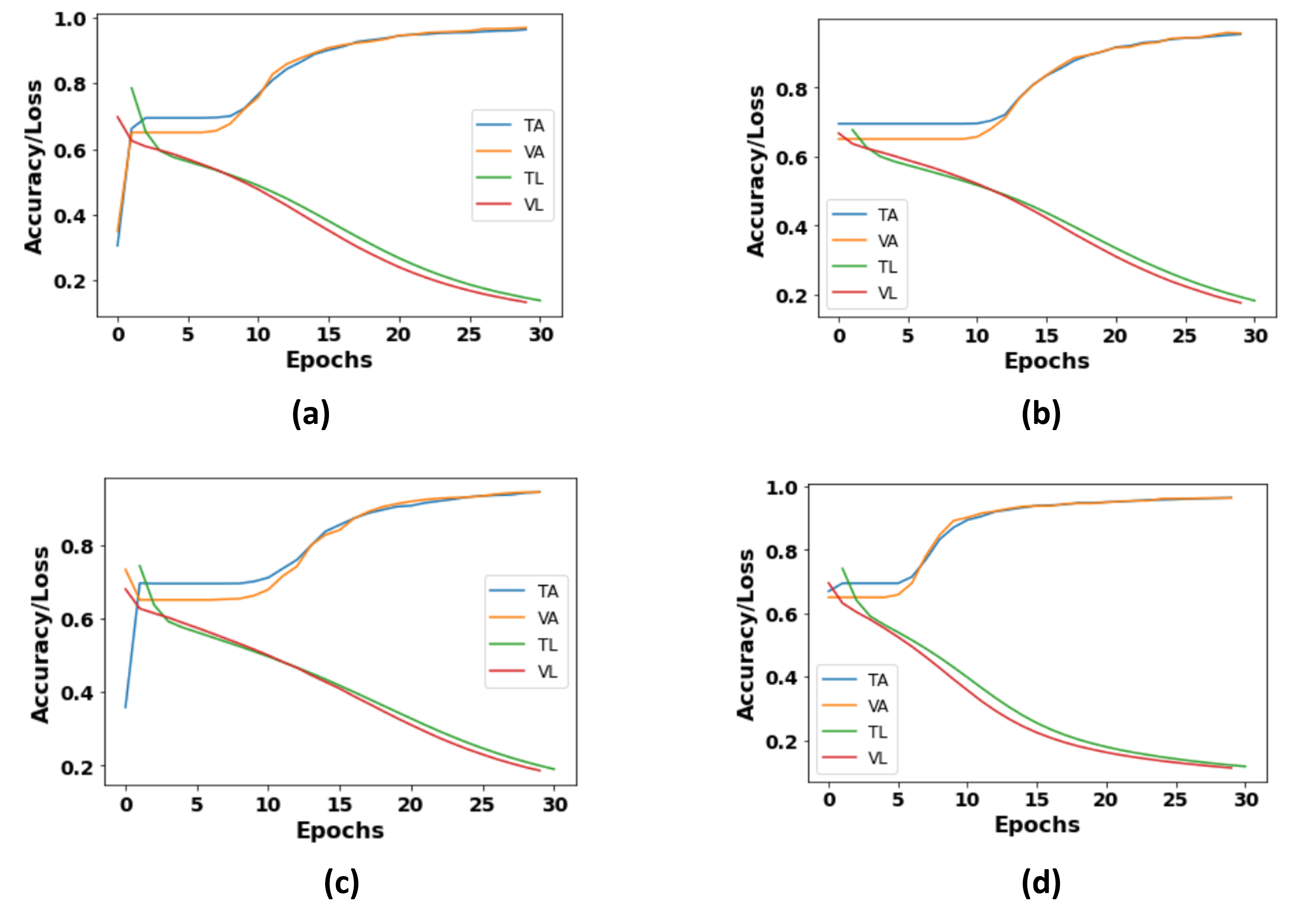}
    \caption{Training and validation performance throughout the training of (a) modified VGG16, (b) ROIN, (c) ROIHEN, and (d) ROIHEDEN during Study Two. TA -- training accuracy; VA -- validation accuracy; TL -- training loss; VL -- validation loss.}
    \label{fig:tac2}
\end{figure*}

In \textbf{Figure}~\ref{fig:tac2}, the performance of modified VGG16, ROIN, ROIHEN, and ROIHEDEN models during the training phase is compared for Study Two. The results show that all of the proposed models continuously improved train accuracy and reduced train loss until epoch 30. 

\subsection{Computational Complexity}
The computational complexity of the modified proposed model was evaluated based on two metrics, namely the number of floating-point operations per second (FLOPs) and the number of parameters (NP) in the model's architecture. \textbf{Table}~\ref{tab:comp} presents the comparative computational analysis between the proposed modified VGG16 model and other transfer learning (TL) based models. The results demonstrate that the modified VGG16 model exhibits lower FLOPs and NP compared to other TL-based models, indicating its superior computational efficiency. Besides the image processing aspect, the TL architecture of the proposed model is similar to modified VGG16 models for other algorithms, such as ROIN, ROIHEN, and ROIHEDEN. Hence, the computational complexity for the proposed algorithms is comparable to that of the modified VGG16 models.
\begin{table}[]
\centering
\caption{Comparison of computational complexity between the proposed model and other existing Transfer Learning-based Models.}
\label{tab:comp}
\resizebox{.5\textwidth}{!}{%
\begin{tabular}{@{}lll@{}}
\toprule
\textbf{Algorithm} & \textbf{FLOPS (Millions)} & \textbf{NP} \\ \midrule
VGG16 & 30960M & 138M \\
ResNet50 & 7751M & 23.58M \\
ResNet101 & 15195M & 42.65M \\
VGG19 & 39037.83M & 20.02M \\
InceptionResNetV2 & 26382M & 55.87M \\
\textbf{Modified VGG16} & \textbf{30713M} & \textbf{15M} \\ \bottomrule
\end{tabular}%
}
\end{table}

\subsection{Models Explainability}
\begin{figure*}[ht]
    \centering
    \includegraphics[width=.7\textwidth]{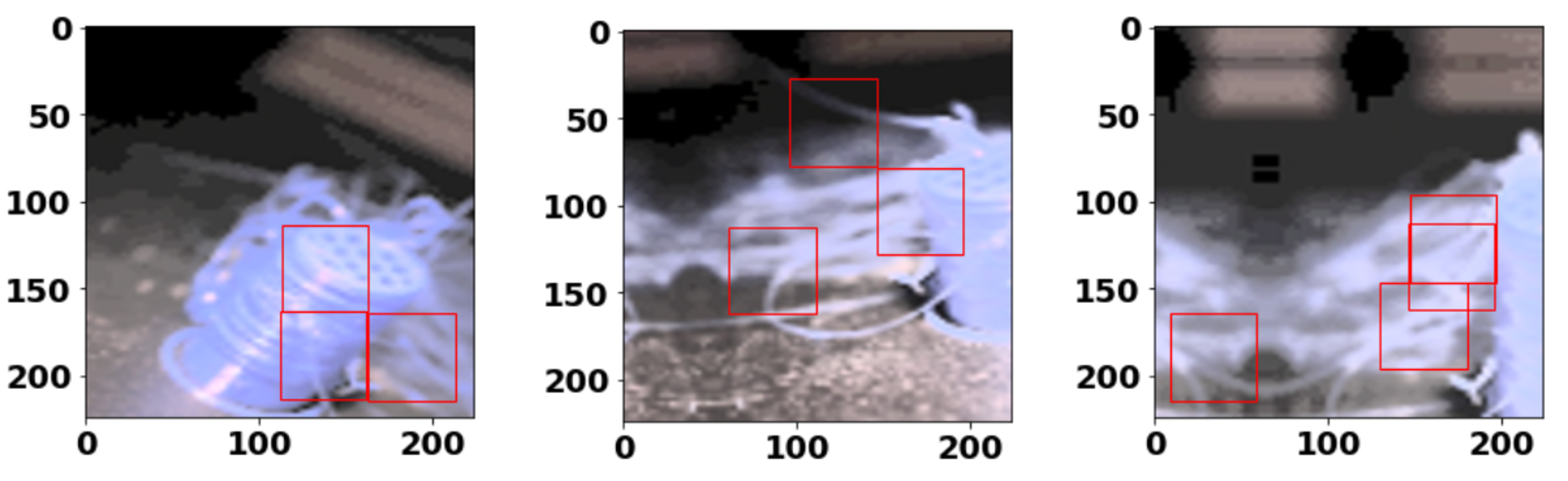}
    \caption{Proposed ROIN model's ability to locate the cylinder and its defect regions.}
    \label{fig:bbox}
\end{figure*}

\textbf{Figure}~\ref{fig:bbox} illustrates the result of using a modified ROIN model to detect defects in a 3D-printed object. The proposed ROIN model has successfully identified the defect in the 3D-printed cylinder object. The red square box in the image indicates the location of the defect, which the model has correctly identified. This is a significant achievement as identifying defects in 3D printed objects is challenging, and accurate detection is crucial for quality control. 

\begin{figure*}[ht]
    \centering
    \includegraphics[width=.7\textwidth]{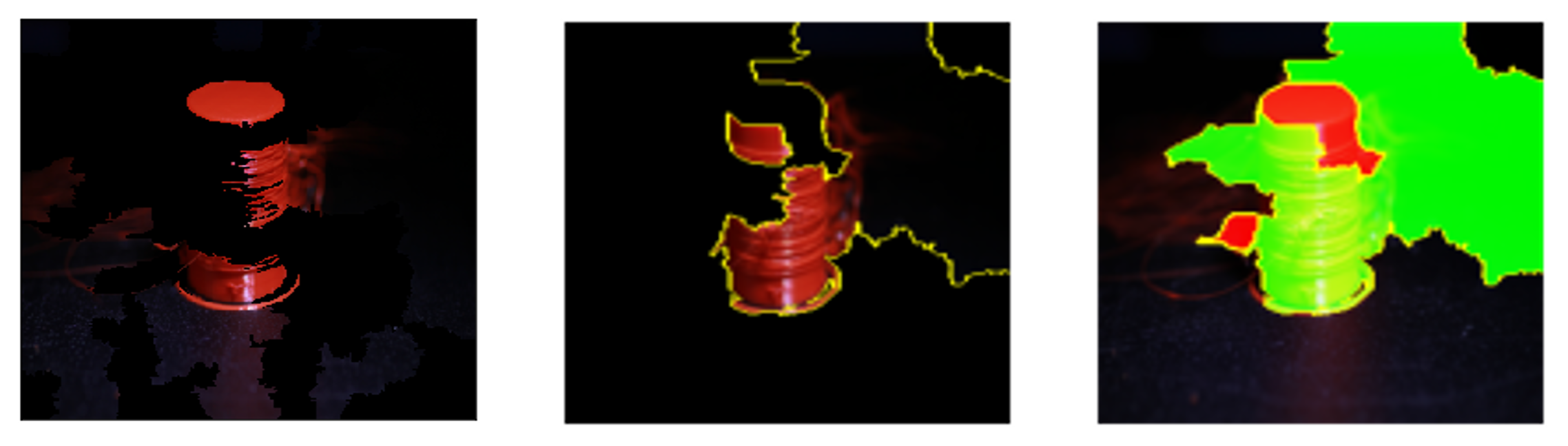}
    \caption{Proposed ROIN models prediction interpretation using LIME.}
    \label{fig:roib}
\end{figure*}

\begin{figure*}[ht]
    \centering
    \includegraphics[width=.8\textwidth]{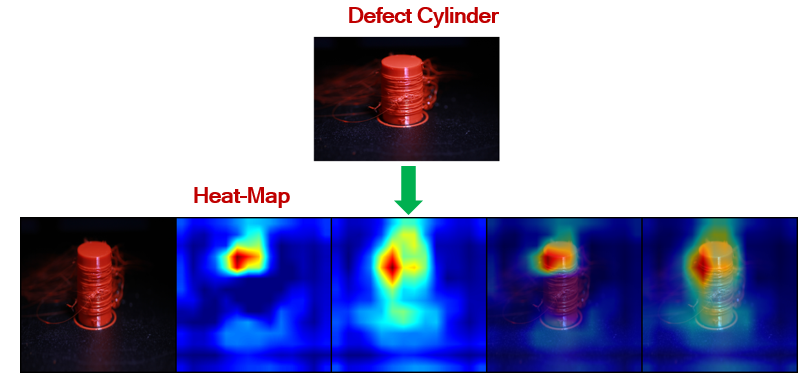}
    \caption{Proposed ROIN models prediction interpretation using Grad-CAM.}
    \label{fig:grad}
\end{figure*}

The LIME approach was used to interpret the predictions made by the ROIN model for the given image, as shown in \textbf{Figure}~\ref{fig:roib}. From the figure, it can be observed that LIME was able to highlight the regions around the defect as being the most important for the model's prediction, which is consistent with the known characteristics of the defect. The heatmap generated by LIME shows a bright red area in the same region as the defect, indicating that this area was most important in the model's decision. 
\textbf{Figure}~\ref{fig:grad} illustrates the Grad-CAM approach used to interpret the predictions made by the ROIN model for the given image. The figure shows that Grad-CAM generated the heatmap highlighting the regions of the input image that contributed the most to the model's prediction. In the case of the given image, Grad-CAM identified the region around the defect as being the most important for the model's prediction, consistent with the findings from LIME. 

\section{Discussion and 
 Overall Findings}\label{discussion}
This study aimed to develop an approach for detecting defects in 3D-printed cylinder images using TL-based CNN models. While the proposed models achieved high accuracy in identifying defects, they faced challenges localizing the specific defect region. The study found that pre-processing steps, such as identifying the defect regions from the cylinder image, are crucial to improving the performance of CNN models. 

The study used the LIME and Grad-CAM approaches to interpret the proposed models' predictions. These techniques enabled identifying essential features in an input image that contributed to the model's prediction. The heatmaps generated from these techniques provided insight into how the model arrived at its decision, allowing for an understanding of its inner workings.

The proposed approaches, including ROIN, ROIHEN, and ROIHEDEN, achieved promising results in detecting defects in 3D-printed cylinder images. The models tested achieved high accuracy, precision, recall, and F1-score values, with all models achieving perfect precision, recall, and F1-score values. The sensitivity of all the models was also 1, indicating that all models correctly classified the defect region as positive. The specificity of all the models was high, with values ranging from 0.8750 to 1.

The results demonstrate that the proposed approaches can improve the performance of CNN models in detecting defects in 3D-printed cylinder images. Furthermore, identifying essential features in the input image using LIME and Grad-CAM approaches can be used to improve the accuracy of the models and reduce the cost and time required for additive manufacturing.

\section{Conclusion and Future Works}~\label{con}
The study proposed three CNN-based approaches, ROIN, ROIHEN, and ROIHEDEN, integrated with three pre-processing steps, ROI, HE, and DE, for detecting defects in 3D printed cylinder images. This research addresses the challenge of identifying defect regions by incorporating additional pre-processing steps. The experimental results demonstrate the effectiveness of these approaches in detecting defect regions within the images.

Additionally, the interpretations of the model's predictions were facilitated by two explainable AI methodologies, LIME and Grad-CAM, which produced heatmaps to shed light on the models' decision-making processes during predictions. The application of LIME and Grad-CAM methodologies has provided significant insights into the functioning of the models, contributing to the overall interpretability and explainability of the proposed approaches. However, future work is required to assess the generalizability of these approaches on a broader dataset and to investigate their capability in detecting different types of defects.

Expanding on this, future research avenues involve further validating the proposed CNN-based approaches using a more extensive collection of data, which could enhance the robustness and applicability of these methods in additive manufacturing. The exploration of these approaches' potential in identifying a wider array of defect types stands as a critical next step. The integration of deep learning and computer vision techniques is poised to play a pivotal role in advancing additive manufacturing, particularly in improving defect detection processes. As such, the continued development and refinement of these approaches promise to significantly impact the quality control measures in additive manufacturing, potentially leading to higher precision, efficiency, and reliability in the production process. This would not only ensure the production of defect-free 3D printed objects but also foster innovation and excellence in the field of additive manufacturing.

\bibliographystyle{unsrt}  
\bibliography{main}

\end{document}